# DARK ENERGY STARS

G. Chapline
*LLNL, Livermore, CA 94025, USA*

Event horizons and closed time-like curves cannot exist in the real world for the simple reason that they are inconsistent with quantum mechanics. Following ideas originated by Robert Laughlin, Pawel Mazur, Emil Mottola, David Santiago, and the speaker it is now possible to describe in some detail what happens physically when one approaches and crosses a region of space-time where classical general relativity predicts there should be an infinite red shift surface. This quantum critical physics provides a new perspective on a variety of enigmatic astrophysical phenomena including supernovae explosions, gamma ray bursts, positron emission, and dark matter.,

## 1. INTRODUCTION

The picture of gravitational collapse provided by classical general relativity cannot be physically correct because it conflicts with ordinary quantum mechanics. For example, an event horizon makes it impossible to everywhere synchronize atomic clocks. As an alternative it has been proposed that the vacuum state has off-diagonal order, and that space-time undergoes a continuous phase transition near to where general relativity predicts there should be an event horizon. For example, it is expected that gravitational collapse of objects with masses greater than a few solar masses should lead to the formation of a compact object whose surface corresponds to a quantum critical surface for space-time, and whose interior differs from ordinary space-time only in having a much larger vacuum energy [1]. I call such an object a "*dark energy star*".

The behavior of matter approaching such a quantum critical surface can be surmised from the behavior in the laboratory of real materials near to a quantum critical point. One prediction is that nucleons will decay upon hitting the surface of massive compact objects. Observation of positron emission from the center of our galaxy may be evidence for the quantum critical nature of the surface of the compact object at the center of our galaxy as well as the Georgi-Glashow picture for nucleon decay. It could also be true that ordinary space-time is close to being quantum critical; indeed, we will suggest that this may be the explanation for dark matter.

## 2. QUANTUM MECHANICS AND RELATIVISTIC ASTROPHYSICS

### 2.1. Wrong Turn at Chapel Hill

In the 1950s a consensus was reached, partly as a result of meetings such as famous meeting at Chapel Hill in 1957, that although quantum effects might be important below some very small distance, on any macroscopic scale the predictions of classical general relativity (GR) should be taken seriously. In the summer of 2000 Bob Laughlin and I realized that this cannot possibly be correct. Indeed I am sure it will be a puzzle to future historians of science as to why it took so long to realize this.

The fundamental reason for the tension between quantum mechanics and GR is the lack of a universal time in GR. A number of arguments can be advanced as to why quantum mechanics requires a universal time. The simplest argument is what time does one mean when one writes down Schrodinger's equation? More subtle arguments involve the existence of non-local correlations. These non-local correlations can exist over cosmological distances (as in Wheeler's delayed choice experiment) and require collapse of the wave function to occur over such distances *simultaneously* with the measurement. At a minimum the notion of simultaneity requires a synchronous coordinate system for space-time. It should also be kept in mind that physical synchronization of clocks requires 2-way communication between the clocks.

The validity of ordinary quantum mechanics requires a universal time, which in turn seems to favor space-times where it is possible to introduce a "synchronous coordinate system". In such a coordinate system the off-diagonal components of the metric $g_{0i}$ have been transformed to zero, and there is a universal time coordinate. As it happens though there exist solutions of the Einstein field equations where it is not possible to introduce a synchronous coordinate system.

The most spectacular examples of solutions to the Einstein equations where the classical behavior of space-time is inconsistent with the existence of a universal time are the rotating space-times; the most famous example being the Godel universe. In these cases universal time fails because the classical space-time manifold contains closed time-like curves. Godel thought that this indicated that there was something wrong with the intuitive notion of time itself. However we prefer to view this strange behavior as an example of the failure of classical general relativity on cosmological length scales. (a view shared by Einstein).

### 2.2. Space-time and Superfluidity

If one visualizes the vacuum as analogous to the ground state of a condensed matter system and ordinary matter as analogous to excited states of this system, then it follows that the atoms in the condensed matter system must move without collective rotation. Indeed, in a coordinate system that is comoving with the collective motion of the atoms the contravariant 4-velocity is just u = 1 and u = 0. In a synchronous coordinate system the covariant 4-velocity would then be u = 1 and u = 0. Consequently, in a non-co-moving coordinate system the 3-velocity v would satisfy curl v = 0. This condition suggests that an appropriate model for the vacuum of space-time is a superfluid .

## 2.3. Black Holes and Quantum Mechanics

One thing that is wrong with black holes vis a vie quantum mechanics is the existence of a space-like singularity which destroys quantum information. A more profound difficulty, though, is the presence of an infinite red-shift surface, i.e. an event horizon, whose existence precludes being able to establish a universal time based on sychronization of atomic clocks. A simple way to see this is to try and use GPS-like coordinates to map out space time (as was suggested by John Synge in 1921). In the case of spherical symmetry this would appear to require just 1 satellite, and allows one to introduce coordinates x0 = 1/2 ( t1 - t2) and x1 = 1/2 ( t1 + t2), where t1 and t2 are the transmission and reception times at the satellite. One can show in the case of a Schwarzschild BH that ds^2 = ( 1- r_g/r) ( dx0^2 - dx1^2) . Obviously the GPS scheme with one satellite doesn't work for r < Schwarzschild radius. In the case of a Schwarzschild BH one would need two independent satellites: one outside the event horizon and one inside the event horizon. In addition the reading on the atomic clock inside the horizon doesn't correspond to a universal time.

In the case of extremely large masses the spatial curvature near the event horizon can be very small. Therefore, one might question whether quantum corrections to classical GR can be important under circumstances where at least locally space-time appears to be quite ordinary. The answer was supplied many years ago by David Boulware [2], who showed that quantum Green's functions near to an event horizon have a cusp-like behavior, regardless of the size of the horizon.

## 2.4. A Simple Thought Experiment

A simple thought experiment makes it clear why it is wrong to assume that classical GR is always correct on macroscopic length scales. Imagine that a Bose superfluid is confined in a vertical column. As a result of the increasing pressure in the fluid as a function of depth it could happen that at a certain depth the speed of sound vanishes. . What is noteworthy about this setup is that the behavior of sound waves near to the critical surface is both well defined, and up to a certain distance from the critical surface, qualitatively indistinguishable from the classical behavior of light outside the event horizon of a Schwarzschild black hole.

It follows from the Schrodinger equation for a superfluid that the dispersion relation for small amplitude waves approaching a critical surface in a superfluid will have the form [1]

$$\hbar\omega_q = \sqrt{(\hbar v_s q)^2 + \left(\frac{\hbar^2 q^2}{2m}\right)^2} \quad , \qquad (1)$$

where $v_s$ is the velocity of sound in the superfluid. In the case of a black hole $v_s$ corresponds to $c(z/2R_g)$, where $R_g = 2GM/c^2$ is the Schwarzschild radius. It follows from Eq. (1) that when relativistic particles approach to within a distance

$$z^* = R_g \sqrt{\hbar\omega/2mc^2} \qquad (2)$$

from the critical surface they will begin to behave like non-relativistic particles with mass m.

As developed previously [1] the surface where the velocity of sound vanishes in our superfluid column ic completely analogous to the event horizon of a classical black hole. However, in contrast with the behavior of waves or particles as they cross the event horizon of a classical black hole, the sound waves in our thought experiment would not pass through the critical surface in an uneventful way. There are two effects in particular that will be important for us. First, in accordance with Eq. (1) the frequency of the waves will become a quadratic function of wave number as they approach the critical surface. Secondly above a certain frequency $Q_0$ the waves will become unstable as they cross the critical surface due to 4-point quantum critical interactions.

On the basis the behavior of sound waves in our thought experiment (which coincides with the general behavior of real materials near to a quantum critical point), we surmise that when matter approaches to within a distance $z^*$ from where GR predicts there should be an event horizon surface ordinary elementary particles morph into heavy non-relativistic-like particles. Since particle interactions in the critical region take place at very high energies as viewed form a distance, one might imagine that these interactions have a universal strength. The universal rate for particle decays would be

$$1/\tau = \hbar\omega^2/mc^2 \quad , \qquad (3)$$

where m is a mass on the order of the Planck mass. Because of the $\omega^2$ dependence in Eq (3), the decay rate will be fastest for those constituent particles with the highest intrinsic energies under ordinary circumstances; i.e. the quarks and gluons inside nucleons.

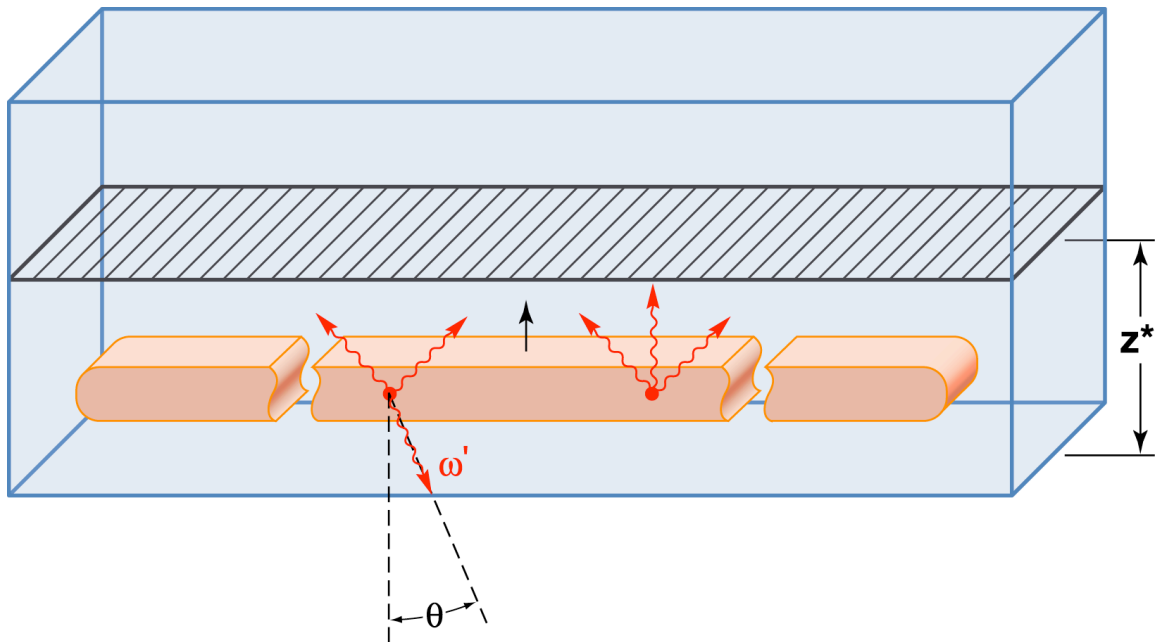

Figure 1: Particle interactions near to the surface of a dark energy star.

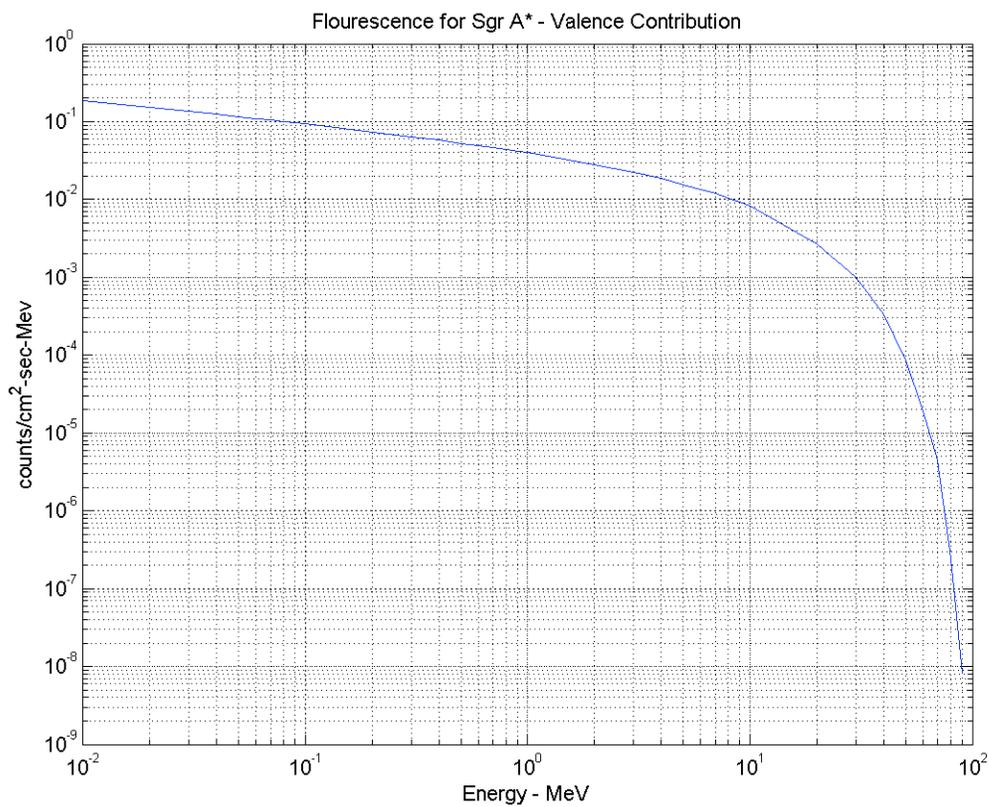

Figure 2: Positron spectrum for $10^6$ solar mass dark energy star.

**Insert PSN Here**

## 3. DARK ENERGY STARS

The new picture that emerges for compact objects is that the interior space-time of the compact object looks like ordinary space-time except that that the vacuum energy is much larger than the cosmological vacuum energy. There is no singularity in the interior. The time dilation factor for the interior metric is positive ( in sharp contrast with the bizarre negative time dilation factor predicted by classical GR), but approaches zero as one approaches the event horizon surface. Near the event horizon classical GR breaks down, and one needs new physics to describe the transition from the interioir to the exterior. From the point of view of GR this transition layer must have unusual properties in order to support large stresses. However, it appears that these objects can be mechanically stable [3].

In the superfluid model the surface of a compact object is a quantum critical shell with thickness $z^*$. When ordinary elementary particles enter this quantum critical region they morph into heavy particles in accord with Eq. (1). In addition they undergo 4-point interactions as shown in Fig. 2. As a result of these 4-point interactions particles can decay; particular particles whose energy before falling into the compact object exceeded $Q_0 = 100$ MeV $\sqrt{(M_o/M)}$, where $M_o$ is a solar mass. The quarks and gluons inside nucleons have energies that exceed this limit for both collapsed stars and the compact objects at the center of galaxies. Thus a signature for the existence of dark energy stars will be nucleon decay [4].

In the Georgi-Glashow grand unified model nucleons can decay; e. g. via a process where a quark decays into a positron and two antiquarks. As it happens an excess of positrons has actually been observed in the vicinity of the center of our galaxy, and there is no conventional explanation for these positrons. Indeed this may be the best evidence to date for dark energy stars

Constituent particles with energies less than the cutoff frequency $Q_0$ will pass through the critical surface, follow diverging geodesics in the interior of a dark energy star, and reemerge through the critical surface. However, for $Q > Q_0$ some of these particles will decay as they pass through the critical surface, leading to decay products directed backwards from the surface of the dark energy star. It's a matter of the geometry of geodesics that this radiation will be beamed backwards in a direction perpendicular to the critical surface and of non-relativistic kinematics that the spectrum of the backward directed radiation has a universal form analogous to the Kurie plot for beta decay [1]. The universal spectrum for positrons from nucleon decay is shown in Fig. 2 in the case of a $10^6$ $M_o$ mass compact object. The Mev energies of these positrons is consistent with the observed spatial distribution of 511 keV annihilation radiation in the galactic bulge. It is interesting that the positron energy spectrum shown in Fig.2 is also qualitatively similar to the spectrum of gamma rays seen in many gamma ray bursts, This suggests that gamma ray bursts may have their origin in matter falling onto the surface of a dark energy star. In the Georgi-Glashow model these gamma rays would not produced directly from nucleon decays, but indirectly from interactions of the decay products during their escape.

Many other aspects of dark energy stars have observable consequences. For example, dark energy stars have a well-defined specific heat, which is quite large; it differs from the usual specific heat of vacuum by a factor on the order of the Planck mass divided by the temperature. This leads to virial temperatures in the infrared, as opposed to a temperature leading to x-ray emission in the case of neutron stars. Thus matter falling onto massive dark energy stars could lead infrared transients.

## 4. DARK MATTER

The point mentioned earlier that the event horizon surface in classical GR can occur in a region of space-time that is nearly flat leads to the question as to whether our local space-time could be near to a quantum critical instability. We will turn this question on its head by suggesting that the Robertson_Walker space-time with which we are familiar may indeed be near to such an instability. In fact, a signal that ordinary materials are close to a critical point is that many small regions of the other phase appear. Applied to our local space-time, this might be interpreted to mean that we should be surrounded by a large number of primordial dark energy stars. Is this the cosmological dark matter?

### Acknowledgments

The results described in this talk were obtained in collaboration with Jim Barbieri, Bob Laughlin, and David Santiago. The author is also very grateful for discussions with Pawel Mazur, Emil Mottola,

This work was performed (in part) under the auspices of the U.S. Department of Energy by University of California Lawrence Livermore National Laboratory under contract No. W-7405-Eng-48.